# Electronic dimensionality of UTe$_2$

L. Zhang[†1], C. Guo[†1], D. Graf[2], C. Putzke[1], M. M. Bordelon[3], E. D. Bauer[3], S. M. Thomas[3], F. Ronning[3], P. F. S. Rosa[3], and P. J. W. Moll[†1]

[1]Max Planck Institute for the Structure and Dynamics of Matter, Luruper Chaussee 149, 22761 Hamburg, Germany

[2]National High Magnetic Field Laboratory, Tallahassee, FL 32310, USA

[3]Los Alamos National Laboratory, Los Alamos, NM 87545, USA

[†]Corresponding authors: ling.zhang@mpsd.mpg.de; chunyu.guo@mpsd.mpg.de; philip.moll@mpsd.mpg.de.

**Superconductivity in the heavy-fermion metal UTe$_2$ survives the application of very high magnetic fields, presenting both an intriguing puzzle and an experimental challenge. The strong, non-perturbative influence of the magnetic field complicates the determination of superconducting order parameters in the high-field phases. Here, we report electronic transport anisotropy measurements in precisely aligned microbars in magnetic fields to 45 T applied along the *b*-axis. Our results reveal a highly directional vortex pinning force in the field-reinforced phase. The critical current is significantly suppressed for currents along the *c* direction, whereas the flux-flow voltage is reduced with slight angular misalignments—hallmarks of vortex lock-in transitions typically seen in quasi-2D superconductors like cuprates and pnictides. These findings challenge the assumption of nearly isotropic charge transport in UTe$_2$ and point to enhanced two-dimensionality in the high-field state, consistent with a change in the order parameter. A pair-density-wave-like state at high fields could naturally induce a layered modulation of the superfluid density, forming planar structures that confine vortices and guide their sliding in the flux-flow regime.**

Heavy-fermion superconductivity in UTe$_2$[1–4] presents outstanding challenges to our understanding of strongly correlated materials hosting $f$ electrons. The main open questions evolve around the remarkably rich field-temperature phase diagram that features three distinct superconducting phases (SC1-3)[5–7] including remarkable reentrant behavior at high fields. At zero and low fields, the system shows prototypical heavy-fermion behavior and superconducting properties consistent with a single-component order parameter (SC1)[1,2,8]. For fields close to the crystallographic *b* direction, a transition into a distinct high-field state (SC2) occurs. Nuclear magnetic resonance and specific heat measurements indicate that this transition is likely accompanied by a change in the order parameter symmetry[9,10]. At higher fields, a uranium-driven metamagnetic transition truncates the superconducting state in a first-order-like transition at 35 T. Lastly, and most enigmatic, is a fully reentrant phase in a small field-angle region around 25°- 40° off the *b* towards the *c* direction (SC3), whose relation to the other phases remains highly contentious[11–13,4].

These enormous critical fields well exceed the Pauli limit[2], which led to proposals of topological odd-parity superconductivity and a surge in experiments investigating the symmetry of superconducting order parameters[14–16,8]. While superconducting states with resilience against high magnetic fields point to unconventional odd-parity states, the tolerance to orbital limiting effects[2] deserves equal attention given its low transition temperature $T_c \sim 2.1$ K.

The first step towards unraveling an unknown superconductor is to establish its phenomenology by determining its Ginzburg-Landau parameters such as the coherence length, which in an orthorhombic system depends on the spatial directions, $\xi_i \{i = a, b, c\}$. Their ratios encode the anisotropy of the superfluid, which is a key parameter to the development of microscopic theories. In addition, reduced dimensionality and fluctuation phase space are key to the physics of cuprates[17], pnictides[18], ruthenates[19], organics[20], and many heavy fermion systems such as CeCoIn$_5$[21], distinct from isotropic unconventional superconductors[22]. The coherence length describes the extent of the vortex core and its anisotropy sheds light on the vortex core deformation resulting from the superfluid anisotropy.

Conventionally, the direction-dependent coherence lengths $\xi_i$ are straightforwardly estimated from the orbitally limited upper critical fields along different crystallographic orientations, as $H_{c2}^i = \frac{\Phi_0}{2\pi\xi_j\xi_k}$ $\{i,j,k = a,b,c\}$, where $\Phi_0 = \frac{h}{2e}$ is the magnetic flux quantum. This logic, however, fails for the SC2 and SC3 phases in UTe$_2$ due to the non-perturbative nature of the magnetic field, which renders the superfluid and its properties strongly field direction dependent, i.e. $\xi_i(\boldsymbol{H})$. The dual role of the magnetic field simultaneously inducing orbital effects as well as tuning and transforming the superfluid itself precludes a crisp identification of the superfluid anisotropy. For example, SC2 only exists for magnetic fields applied close to the $b$ direction[23], and critical fields of SC2 along other directions cannot meaningfully be defined or determined. Still, as long as SC2 can be described by a thermodynamic expansion as in Ginzburg-Landau theories, at every point in field a well-defined set of $\xi_i$ and their anisotropies exists. A second complication is the experimental identification of orbitally limiting fields, as other field-tuned effects may suppress superconductivity. SC2 for fields along the $b$-direction is abruptly terminated by a metamagnetic transition well before a putative orbital limit occurs at unknown higher fields. Despite the importance of the field evolution of the superfluid anisotropy in our understanding of UTe$_2$, its investigation is a challenge due to the temperature and field scales of its superconductivity and the non-trivial role of the magnetic field.

This study aims to assess the anisotropy of the high-field phase SC2 by probing the electronic transport anisotropy in the flux-flow regime via critical currents. The anisotropic voltage response for electric currents running in narrow bars along different crystallographic directions allows an estimate of the degree of anisotropy without comparing different field directions. The main finding is a textured, quasi-2D superconducting state in SC2, in contrast to the weakly anisotropic superconductivity at low fields (SC1). The experiments are based on crystalline microstructures of UTe$_2$ carved by focused ion beam[24] from ultraclean crystals grown by the molten salt flux (MSF) method[25].

## Results

The electrical resistivity of UTe$_2$ along all three crystalline directions of this orthorhombic compound is measured in carefully aligned microstructures (Fig.1a, see more details in Methods). In this "L-bar" geometry, two rectangular bars angled at 90° allow two four-probe measurements in series. Reducing the bar cross-section to the micron-scale is key to achieving the high current densities required to enter the flux-flow regime well below $H_{c2}$, which will provide additional, non-linear insights into transport complementing previous magnetoresistance studies on bulk crystals at a low current density[5–7]. Two microstructures, S1 featuring $a$- and $c$-channels and S2 featuring $b$- and $c$-channels, respectively, have been studied to provide a complete picture of the anisotropy.

Focusing first on the normal state resistivity (Fig.1b,c), $\rho_a$ and $\rho_b$ increase with decreasing temperature until reaching their maximum value at 60K and 90K. This typical behavior in Kondo lattices has been associated with the onset of Kondo coherence[26]. The out-of-plane resistivity, $\rho_c$, is markedly different: on cooling, it shows metallic behavior at high temperatures until 60 K, followed by a sharp peak around $T^* = 15.5$ K well below the Kondo scale of $\rho_a$ and $\rho_b$, but is close to the Kondo coherence temperature estimated by STM measurements[14]. This sharp peak in $\rho_c$ has been associated with magnetic fluctuations[26] or a much lower Kondo coherence temperature due to orbital-selective Kondo scattering[27]. At $T_c$~2.05 K (defined by $\rho = 0$), a sharp transition into the noise floor is observed, indicating a robust superconducting state. The sharpness of the transition indicates the absence of strain gradients, whereas subtle device variation of $T_c$ and $H_{c2}$ indicates some homogeneous strain arising from differential thermal contraction (see SI for strain estimations). Given their similar behavior in temperature, not surprisingly $\rho_a/\rho_b$ is relatively temperature-independent with a rather low anisotropy of 2-3 (Fig. 1c). The largest resistance anisotropy at any temperature is found with respect to $\rho_c$. The ratio $\rho_c/\rho_b$ is ~5 at 300 K and remains stable as the temperature decreases until around 20 K, below which it increases rapidly and reaches a factor of 50 just above $T_c$. Overall, both the in-plane and out-of-plane anisotropies are compatible with a layered conductor not quite in a quasi-2D limit, which is consistent with the current Fermi surface models of strongly corrugated cylinders[28] (see SI for further discussion).

Our results agree reasonably well with a previous resistivity anisotropy study[26], yet remarkable and vital differences exist. First, we consistently observe the lowest resistance at all temperatures in the $b$ direction, perpendicular to the uranium chains along the $a$ direction, while Ref. 26 observes the lowest resistance along the $a$ direction. As both measurements consistently observe the field-reinforced phase at $H||b$, accidental misalignment can be safely excluded. At room temperature, the resistivity magnitudes in bulk crystal measurements in zero field are in quantitative agreement with the microbars from which they were obtained and exclude potential fabrication-related issues (see Fig. S2). The quality of the structures is further evidenced by high residual resistance ratios (RRR = $\rho(300K)/\rho(0K)$) along $a$ and $b$ axes (RRR$|_{\rho_a} \approx 1000$, RRR$|_{\rho_b} \approx 350$), as well as the sharpness of the transitions and the clear presence of quantum oscillations. Second, Ref. 26 observes a remarkably low value for $\rho_c$, which even falls below $\rho_b$ at intermediate temperatures. This led to the key interpretation of the normal state as a rather isotropic conductor, which contradicts electronic structure models that propose only a corrugated cylindrical Fermi surface along the *c* direction. As a result of this observed isotropy, the presence of a small, closed 3D Fermi surface is proposed. The existence of such a pocket is under active debate, with quantum oscillation evidence for it only observed in tunnel diode oscillator measurements, being challenged by its absence in electrical resistivity and ARPES measurements[29]. Our results paint the picture of a much more anisotropic normal state in the MSF crystals, which is further supported by the observation of strong vortex lock-in in the high field phase discussed below. Our transport anisotropy is well explained by a corrugated cylinder alone, suggesting that UTe$_2$ is a low-dimensional metal. A possible route to reconcile these contradicting experiments would be a non-trivial role of chemistry given that Ref. 26 employed chemical-vapor-transport-grown crystals that show a lower superconducting transition temperature ($T_c$ = 1.6 K) and smaller residual resistivity ratios (RRR ~ 20).

The magnetoresistance of the normal state shows a non-trivial field dependence from a competition of orbital effects and magnetic tuning of the 5*f* electrons. As no exotic high-field behavior or reentrance is observed for fields along the *c* direction, here one may expect a more conventional orbitally limited upper critical field ($H_{c2}$) that may be determined from the magnetoresistance in the $\rho_a$ and $\rho_c$ channels (Fig.1d). The channels show a difference (2.6T) of upper critical field of the SC1 phase, consistent with their small difference in $T_c$ likely rooted in the non-zero thermal differential contraction. When $H > H_{c2}$, $\rho_a$ shows a $B^2$ dependent increase, typical for the transverse magnetoresistance of a metal. Clear Shubnikov-de Haas (SdH) oscillations appear on top of this background (Figs.1e,f), evidence for the high sample quality and homogeneity in the FIB microstructures. The oscillation frequencies and their angle dispersion are consistent with previous quantum oscillation results supporting a quasi-2D fermiology of UTe$_2$[28,30](See SI for comparison), and do not show any sign of small 3D FS pockets. Naturally, this experiment reveals only a few of the lightest orbits, given the relatively high sample temperature in the 45 T magnet ($T \geq 380$ mK) and the heavy masses in UTe$_2$. In contrast, $\rho_c$ first increases up to a shallow maximum around $H \sim 26T$, followed by a small decrease. Such behavior is commonly seen in the longitudinal magnetoresistance of magnetic materials and attributed to a suppression of spin-scattering with polarization of moments in the absence of orbital effects.

Overall, a consistent picture of an anisotropic normal state emerges, which is in good agreement with the moderate anisotropy of the low-field superconductivity SC1. $H_{c2} \parallel a$ = 8 T and $H_{c2} \parallel c$ = 15 T may be easily obtained as the boundary between SC1 and a paramagnetic metallic state from previous reports[7], yet $H_{c2} \parallel b$ of SC1 is difficult to access as the nature of the SC1-SC2 transition remains to be clarified. One might be tempted to assume $H_{c2} \parallel b$ = 20 T, following the zero-temperature extrapolation of the heat capacity anomaly[9] in line with ac susceptibility measurements[10,31]. Under these assumptions, one estimates the Ginzburg-Landau coherence lengths of SC1 at zero temperature as $\xi_a = 3.3nm, \xi_b = 8.4nm, \xi_c = 6.3nm$ from $H_{c2}^x = \phi_0/2\pi\xi_y\xi_z$. Of course, these numbers are rough estimates at best, given how field-dependent the superconductivity itself is found to be. Even in the rather conventional $H \parallel c$ direction without reentrance, a clear deviation from the conventional Werthamer-Helfand-Hohenberg behavior describing limiting through orbital effects is observed, which challenges the interpretation of the measured $H_{c2}$ as an orbital limit field[2]. This directly reflects the difficulty of assessing anisotropy in such field-tuned systems. Importantly, for the following vortex lock-in discussion, these numbers greatly exceed the unit cell size in any direction, with a comfortable safety margin as to the uncertainty of the orbital limiting fields. Not surprisingly, SC1 is expected to host Abrikosov vortices as a moderately anisotropic yet three-dimensional superconductor.

Next, we examine the field-reinforced SC2 phase for fields along the *b* direction in which the previous analysis of upper critical fields cannot be applied. In this state, $\rho_a$ and $\rho_c$ measured at low current densities ($J < 0.019 kA/cm^2$) display distinct field dependencies (Fig.2). $\rho_a$ shows zero resistivity up to the metamagnetic transition at $H_m$ = 35 T. Note the field zero offset of 11.5 T, the minimal field of the hybrid magnet system at the NHMFL in Tallahassee, below which robust zero resistance has been observed under all conditions in a superconducting magnet (see Fig.S3). Given the high vortex density of $1.7 \times 10^4 \ \mu m^{-2}$ at these fields, the zero-resistance state indicates a relevant region in phase space characterized by well-pinned vortices. In contrast, a large signal in $\rho_c$ emerges at the same current density. The voltage response reaches levels in the same order as the normal state resistance in zero field, 0.1 mΩ·cm, indicating at substantial flow of vortices. This highly anisotropic response to rather low current densities already suggests a significantly anisotropic superconducting state in SC2.

To further investigate the nonlinearity in the flux-flow state (Fig.2c-e), a direct current (dc) $I_{dc}$ offset was superimposed on a small alternating current (ac) $I_{ac} = 2 \ \mu A$ (*a*-channel: 0.019 $kA/cm^2$, *c*-channel 0.014 $kA/cm^2$, reflecting their slightly different geometric factors). The strong field dependence of the ac-resistance $R_{ac}^c = V_{ac}/I_{ac}$ for $J \parallel c$ reveals rich vortex physics as the system is tuned by the magnetic field. Starting from a robust zero-resistance state at low fields, flux-flow voltage grows as the boundary between SC1 and SC2 is approached.

This initial region is characteristic for an approaching irreversibility line[32], with a gradual softening of the critical currents. This is followed by a narrow region of quasi-ohmic behavior characteristic of free vortex flow as in a Bardeen-Stephen picture[33], which is indicated by a similar AC voltage for all applied biases. Such typical transitions between vortex solids and liquids mark the irreversibility line crossing into a free flow of vortices within SC1. The collapse of the hysteresis of magnetostriction experiments suggests a vortex liquid phase between 15 T and 22 T[9], which quantitatively matches our transport data (shaded region in Fig.2). Further signatures for an irreversibility line at this field boundary have been reported in ac-susceptibility and the $a$ direction resistance, $\rho_a$, from which an intermediate vortex liquid phase in-between SC1 and SC2 has been deduced[31,34]. Beyond 22 T, the curves deviate again, indicating the non-ohmic response of finite pinning within SC2. Further increasing the field suppresses the voltage at equal drive. Considering the behavior of $T_c$ being enhanced with increasing field in SC2 phase, a natural explanation for this result is the $\delta T_c$ type of pinning[35], which describes the pinning force induced by disorder in $T_c$ and scales with $(1 - T/T_c)^{-1/2}$. Interestingly, closer inspection of the transition region under finite bias exhibits a sharp step at $H^* \sim 21.7$T, delineating the quasi-ohmic from the activated transport region ($H > H^*$). The step is in close proximity to the phase boundary SC1-SC2 observed in specific heat[9] at $H \approx 19$ T when $T = 0.3$ K, and marks a clear signature of the transition in the vortex response. Recent suggestions of an order-parameter symmetry change are compatible with such an instantaneous response in the vortex matter[36]. The occurrence of a discontinuous response may well be a result of the unconventional situation of approaching a vortex irreversibility line from a vortex solid state at high fields, a situation unique to field-driven SC-SC transitions. Indeed, the step-like transition moves to lower fields under increasing bias currents. This is opposite to expectations from self-heating effects, as the irreversibility line on the high field side has been found to move to higher fields with higher temperatures due to the field-reinforcement of SC2[9]. It will be interesting to explore the hypothesis further that the SC1-SC2 transition could be shifted by bias currents as a response to the different phase stiffness in these distinct superconducting condensates.

The situation for current along the $a$ direction, however, is clearly different. $\rho_a$ remains at low-resistive conditions up to the highest applied current densities of $0.44$ kA/cm$^2$, indicative of a well-pinned vortex system. A weak suppression of flux flow resistance is found around 22 T, which matches with the $H^*$ step in the $c$-direction measurements. Together, this implies an increase in vortex mobility at $H^*$ when driven by $c$-direction currents and suppression when driven by $a$-direction ones, with a substantially lower critical current for the $c$ direction.

Such mobile vortex matter of SC2 is only observable for fields well aligned along the $b$ direction (Fig.3). Its flux-flow resistivity $\rho_c$ is highly angle dependent, terminating in a sharp drop when the field is tilted away from the $b$-axis (Fig.3b). Over most of SC2, the flux flow voltage vanishes within 4° of misalignment. The peaks are well described by Lorentzian functions of form $\rho = \rho_0 + \frac{2A}{\pi} \frac{w}{4(x-x_c)^2 + w^2}$ and the peak width $w$ shows nonmonotonic field dependence (Fig.3c). Starting out as rather broad peaks at 22T, increasing the field narrows the response down to a minimal peak width of 2° at 28 T. Further increasing the field broadens the peaks, while the overall flux flow voltage at $H||b$ continues to decrease. This increase is reminiscent of the growth of the quadratic temperature coefficient of resistance in a recent Kadowaki-Woods analysis[37], suggestive of a fluctuation-driven weakening of the superconducting coherence as the metamagnetic transition is approached.

The $a$-axis flux flow reacts to a similar angle range (Fig.4). While $\rho_a$ maintains at zero-resistance at the low probing currents for fields along $b$ ($\theta = 0°$), dissipative behavior onsets at small angles ($\theta > 4°$). A notable dissipative band emerges around 20 T, reflecting the vortex liquid phase also observed in $\rho_c$. Within SC2, $\rho_a$ shows a robust zero-resistance state up to high angles above 10°, which likely is not due to vortex anisotropy but a result of the weakening order parameter as the system approaches the boundary of the field-reinforced phase upon field rotation. Accordingly, it returns to the normal state value at around 20°.

## Discussion

The flux-flow dynamics probed by both current channels are significantly dependent on the vortices' alignment with the $b$-axis of the crystal. Such sharp lock-in effects are commonly observed in the intrinsic pinning regime of layered superconductors such as pnictides, cuprates, or organics[38–40]. In this regime, the vortex cores are confined to reside between conductive planes of high superfluid density which lowers the condensation energy cost of the normal core region (Fig.3a). A highly directional barrier results, with easy vortex motion along the planes, driven by currents out-of-plane, and strong pinning for vortex motion perpendicular to the planes, driven by in-plane currents. A hallmark of intrinsic pinning is its sharp dependence on the field angle. At a small out-of-plane angle, the vortex system transitions and realigns with the external field by crossing through the planes, inducing so-called "pancake vortices". The directional pinning is lost at that point, and in particular, the free flow of vortices parallel to the planes stops, as seen here in UTe$_2$. Acceptance angles around 5° are typical in structurally layered, highly anisotropic superconductors such as iron-based SCs[38], organic SCs[39], and cuprates[40]. Hence, the observation of sharp intrinsic pinning suggests a similar degree of low dimensionality in UTe$_2$ and these classes of layered unconventional superconductors, under the implicit assumption that a 4° field rotation does not fundamentally alter the order parameter within SC2. Indeed this is in line with a quasi-2D Fermi surface shown by quantum oscillations and ARPES experiments[28–30] and the resulting normal state anisotropy suggests to consider UTe$_2$ as a quasi-2D superconductor.

Interestingly, this anisotropy in SC2 differs from SC1, where the previous estimate suggested the shortest coherence length in the *a*, not *c* direction. Either SC1 cannot be interpreted as an orbitally limited phase with field-independent Ginzburg-Landau parameters in any direction, or the transition into SC2 for $H \parallel b$ substantially changes the magnitude and direction of the anisotropy, or both.

It is important to differentiate between anisotropic and low-dimensional superconductors. Superconductors are anisotropic when their coherence lengths depend on the spatial direction, while they are quasi-2D when the coherence length in one direction falls below the interlayer distance, typically given by the crystalline unit cell. Even highly anisotropic superconductors are often three-dimensional when their shortest coherence length still exceeds any intra-unit-cell modulation of the order parameter. For a vortex to resolve such intrinsic superfluid density modulations in a quasi-2D superconductor and to lower its potential energy by being locked in between the planes, its core must be smaller than the interlayer distance. Conversely, lock-in phenomena are observed when the vortex core diameter in the out-of-plane direction, $\sim 2\xi_c$, is compatible with the interlayer separation, $d_c \geq 2\xi_c$[38]. This is not an easy criterion to fulfill in UTe$_2$. The upper bound for a coherence length compatible with intrinsic pinning, $\xi_c = \frac{d_c}{2} = 0.7\ nm$, indicates a remarkably small value, similar to that observed in high-T$_c$ materials with massive upper critical fields. If one keeps $\xi_a \sim 3.2\ nm$ as estimated from the critical fields in SC1, one finds a lower bound of $H_{c2}^{b,min} = 145$ T suggests that SC2 might extend to extreme fields if it were not truncated by the metamagnetic transition. Another possibility is that $\xi_a$ substantially grows within SC2, thus lowering $H_{c2}$. In this scenario, the lowest possible value for $\xi_a \sim 14\ nm$ is found if the orbital limit coincides with the metamagnetic transition, $H_{c2}^b = 34$ T.

An alternative solution, however, is possible. The vortices sense a superfluid density modulation compatible with their physical size, which can also be fulfilled if superconducting planes spaced multiple unit cells apart self-assemble. In such a layered pair-density-wave-like scenario (PDW), $\xi_c \sim \frac{l_c}{2}$, where $l_c$ denotes the periodicity of the superconducting modulation along the *c* direction. While our experiment cannot distinguish between these scenarios, the PDW case provides a unified picture of the superconducting phases of UTe$_2$, consistent with several experimental observations. The PDW period may be multiple unit cells, thus relaxing the extreme field and anisotropy scales required by atomic-scale coherence lengths. The apparent change in principal directions of anisotropy at the SC1-SC2 transition may also be understood within this framework. While the estimate of coherence lengths in SC1 is crude, the relative ratios of $H_{c2}$'s are hard to reconcile without $\xi_a$ being the shortest coherence length, while clearly in SC2 the shortest one is $\xi_c$, given the observed lock-in effect. Such a change of anisotropy would naturally come with a change in PDW periodicity. Lastly, this could be seen as support of Lebed's argument for field-induced superconductivity in layered conductors[41]. When superconductivity lowers the energy of a Fermi liquid in layered metals under in-plane fields, a reciprocal force pushes Fermi liquids to self-layer if they have the freedom to do so. In normal metals with bandwidths in the eV, this is clearly impossible due to the prohibitive cost of distortions of the lattice. However, in a U-dimer system, valence-, spin- or multipolar modulations provide low-energy mechanisms for a material to self-layer. Within that scenario, the high-field phase SC3 may be interpreted as a different allowed PDW wave vector along the (011)-diagonal. Despite the absence of a bulk signature[42,43], an instability towards a PDW already of SC1 has been suggested by STM in zero field[44] as well as evidence for its field-modulation[45]. Although these observations can only evidence a surface density-wave state, the possibility of its evolution to a bulk state through the self-layering mechanism mentioned above is possible at high magnetic fields.

Overall, our flux-flow experiments on UTe$_2$ have directly observed substantial anisotropy and real-space texture in the superconductivity of SC2, which follows the electronic anisotropy suggested by normal state transport and the known Fermi surface anisotropy. The nature of the superfluid resembles that of a layered quasi-2D superconductor, which should put constraints on microscopic theories of superconductivity in UTe$_2$. The observation of vortex lock-in *per se* fits well into a picture of a modulated superconducting order parameter, with a one-dimensional modulation along the *c* direction as seen in structurally layered systems such as La$_{2-x}$Ba$_x$CuO$_4$[46]. Uranium compounds, and even elemental uranium, have been host to a wealth of field-induced density wave phenomena[47,48], and highly resilient forms of superconductivity are found once the effective dimensionality is reduced – a common theme in high-T$_c$ materials.

# Method

### Crystal synthesis and basic characterization

Single crystals of UTe$_2$ were grown through a molten salt technique using an equimolar mixture of sodium chloride and potassium chloride (NaCl + KCl) as reported previously[25]. The crystallographic structure of our crystals was verified at room temperature by a Bruker D8 Venture single-crystal x-ray diffractometer equipped with Mo K-$\alpha$ radiation. To ensure that the samples only show a single superconducting transition temperature, specific heat measurements were performed using a Quantum Design calorimeter that utilizes a quasi-adiabatic thermal relaxation technique.

### Microstructure calibration

The microstructures of UTe$_2$ are fabricated by Thermofisher Helios Ga FIB or Hydra. The lamellae of the two samples used were cut from an MSF-grown crystal with a superconducting $T_c = 2.1K$, and then transferred onto a sapphire substrate with Pt welding deposited by FIB. These lamellae were etched with radio frequency (RF) argon plasma to remove surface oxide layer, which was followed by a high-power Au sputtering in the same vacuum chamber to form electric contact. The gold sputtered lamellae were patterned to L-bar geometry and protected with FIB-deposited carbon in Hydra system. The devices were calibrated in Quantum Design PPMS system with 9T superconducting magnet for the temperature dependence of resistivity.

### High-field measurements

High-field transport measurement was performed in the 45T hybrid magnet at the National High Magnetic Field Laboratory. Transport signals were read out via Stanford Research 86x lock-in amplifiers. AC and DC currents are applied via the Stanford Research CS580 voltage-controlled current source.

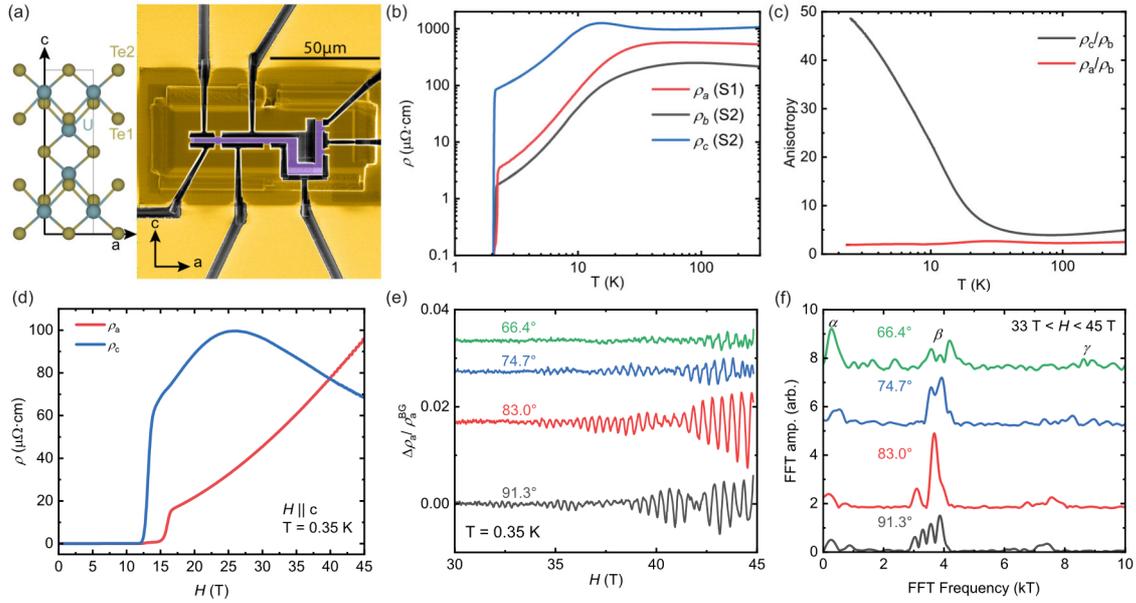

**Fig.1. Basic properties of UTe$_2$. a**. the unit cell of UTe$_2$ (left) and SEM image of a FIB microstructure with channels along the *a*- and *c*-axis (right). **b.** The resistivity along different crystallographic axes from 300K to 2K **c.** Resistivity anisotropy of UTe$_2$ from 300K to 2K, expressed in $\rho_c/\rho_b$ and $\rho_c/\rho_a$. **d.** Magnetoresistivity in 12-45T for the magnetic field along the *c*-axis, with the current along the *a*- and *c*-axis at $T = 0.35\ K$. **e.** Shubnikov-de Haas oscillation when the magnetic field is rotating in *bc* plane, with the angles between *H* and *c*-axis labeled. **f.** FFT spectrum of the Shubnikov-de Haas oscillations, with a field window from 33T to 45T.

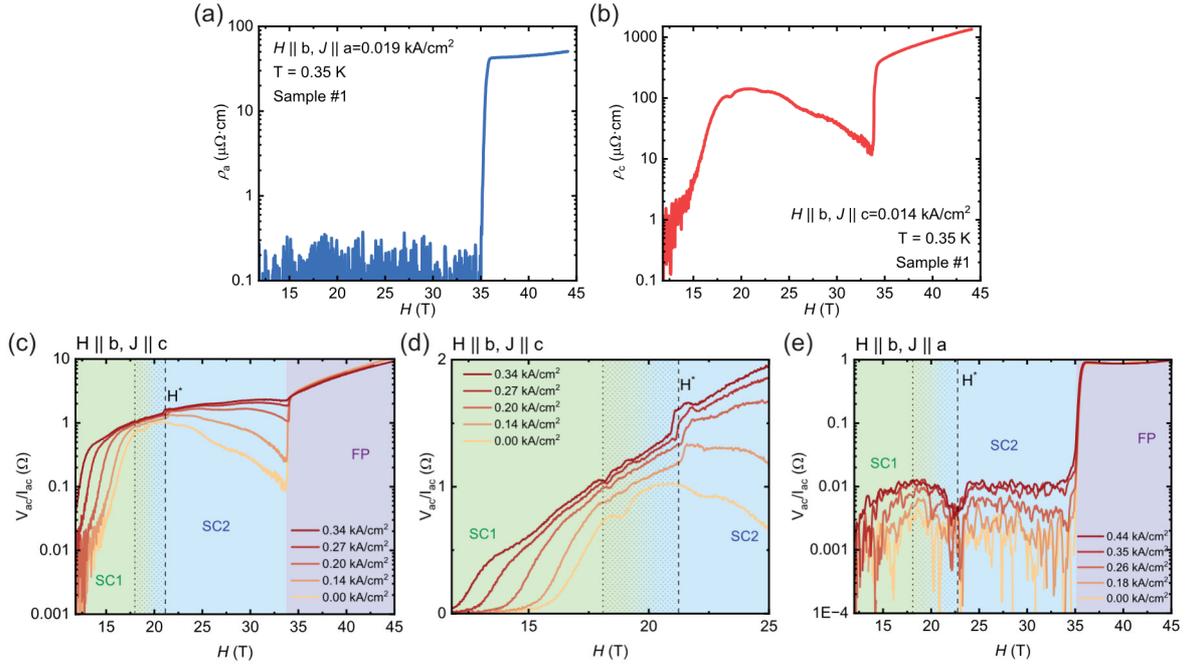

**Fig.2. Anisotropic flux flow in SC2 phase of UTe$_2$. a.** Resistivity $\rho_a$ and **b**. $\rho_c$ when $H||b$, at $T = 0.35K$. Nonlinearity of flux flow signal when $H||b$, with $J||c$, on **c.** logarithmic and **d.** linear-scale. **e.** Nonlinearity for $J||a$.

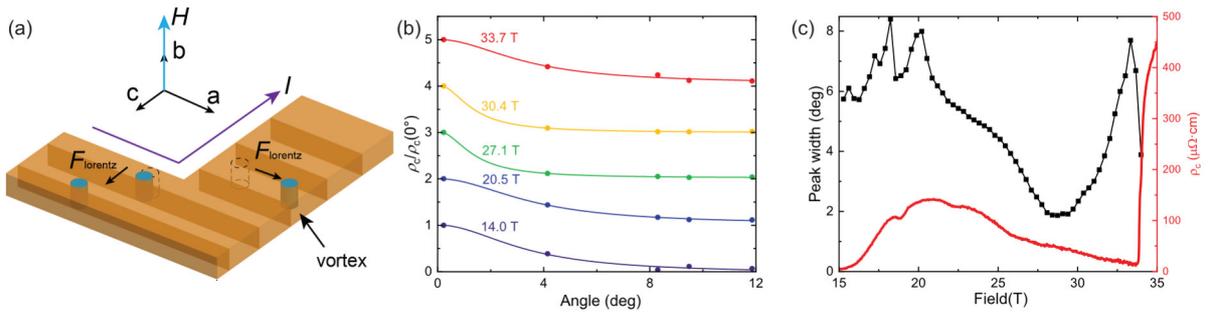

**Fig.3. Layered superconducting texture a**. Sketch of layered model of vortex motion in UTe$_2$. **b**. Angular sweep of $\rho_c$ (normalized by $\rho_c(0°)$) with magnetic field rotated from *b*-axis to *c*-axis (offset by 1 for different fields), solid curves are fit by Lorentzian function. **c**. Peak width $w$ of the Lorentzian fit changing with magnetic field.

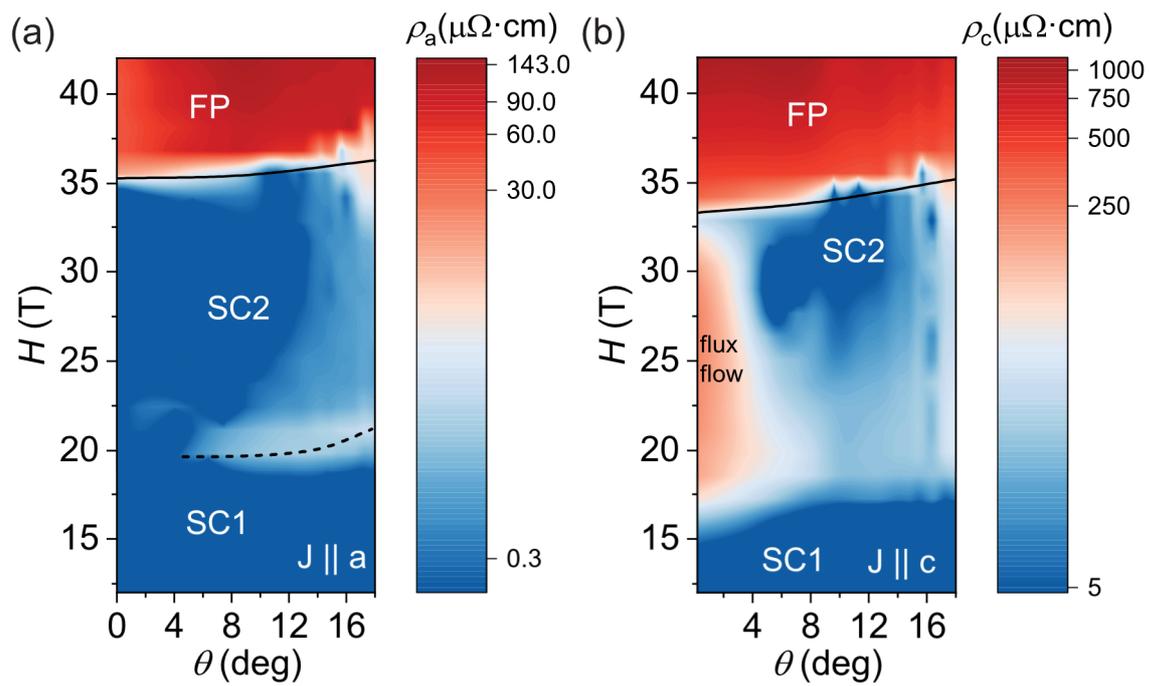

**Fig.4. Comparison of current direction dependence on flux flow voltage**, for. **a.** $J||a$ and **b.** $J||c$. The dashed line labels the dissipative band around 20 T.

**Acknowledgements**

This work was supported by the DOE Office of Basic Energy Sciences, Materials Sciences and Engineering Division project 'Quantum fluctuations in narrow band systems'. C.G. acknowledges financial support by the European Research Council (ERC) under grant Free-Kagome (Grant Agreement No. 101164280). The high field experiments were performed at the National High Magnetic Field Laboratory, which is supported by National Science Foundation Cooperative Agreement No. DMR-2128556 and the State of Florida.

Supplementary information for:

# Electronic dimensionality in UTe$_2$

L. Zhang, C. Guo, D. Graf, C. Putzke, M. M. Bordelon, E. D. Bauer, S. M. Thomas, F. Ronning, P. F. S. Rosa, and P. J. W. Moll1

### A. Strain effect of microstructure

The strain induced by mismatch of thermal expansion coefficient on our microstructure samples can be roughly estimated. The thermal strain of the sapphire substrate we used is about $\epsilon = -0.0008$ at 2 K, while the thermal strain of single crystal UTe$_2$ was measured in previous measurements[1], which gives $\epsilon_a = -0.00345, \epsilon_c = -0.00347$. Therefore, we can estimate that the thermal strain on microstructure when $T = 2K$ is about:

$$\epsilon = \epsilon^{UTe_2} - \epsilon^{sub} = -0.0027$$

The Young's modulus of UTe$_2$ was also measured by other experiments[2], which is $E_a = 90.3\ GPa$ along $a$-axis and $E_c = 95.9\ GPa$ along $c$-axis. The stress on microstructure is:

$$\sigma_a = \epsilon \times E_a = -0.244\ GPa$$

$$\sigma_c = \epsilon \times E_c = -0.259\ GPa$$

The strain effect on $T_c$ of UTe$_2$ was previously studied[3] and shows that for absolute stress from 0~0.4 GPa, $T_c$ shows a linear dependence on $\sigma$, with $\frac{dT_c}{d\sigma_a} = -0.87\ K/GPa, \frac{dT_c}{d\sigma_c} = 0.56 K/GPa$. The change of $T_c$ on our microstructure can then be estimated as:

$$\Delta T_c^{a\ chn} = \sigma_a \frac{dT_c}{d\sigma_a} = 0.212\ \text{K}$$

$$\Delta T_c^{c\ chn} = \sigma_c \frac{dT_c}{d\sigma_c} = -0.145\ \text{K}$$

$$\Delta T_c = \Delta T_c^{a\ chn} - \Delta T_c^{c\ chn} = 0.357\ \text{K}$$

This estimation doesn't consider the distribution of stress on the whole sample, which may lead to an overestimation of the difference of $T_c$ between two channels. In $\rho(T)$ curves, we observed the actual $\Delta T_c$ is about 0.15 K, which shows a good consistency with the above calculation.

B. Shubnikov-de Haas oscillation

Here we compare the main frequencies in the Shubnikov-de Haas oscillations observed in our experiments when $H||c, J||a$ with previous reports in Table S1.

|  | This paper ($\theta = 91.3°$) | Broyles et al. 2023[4] | Eaton et al. 2024[5] | Aoki et al. 2023[6] |
|---|---|---|---|---|
| Type of experiment | 4-point resistance | Tunnel diode oscillator | Magnetic torque and proximity diode oscillator | Field modulation |
| Frequencies | 3.037 kT<br>3.3 kT<br>3.565 kT<br>3.864 kT | 3.2 kT<br>3.7 kT<br>4.1 kT | 3.5 kT | 3.14 kT<br>3.33 kT<br>3.67 kT |

Table. S1 Comparison of quantum oscillation frequencies when $H||c$ in different experiments

From the comparison, it is clear that the quantum oscillation frequencies measured in our transport measurement using FIB-fabricated microstructures are comparable to all other reports. However, our measurements present more details about the peak profile in the frequency range 3~4 kT, featuring four almost evenly spaced peaks with similar amplitudes. These four frequencies are closely located and can merge to a singular peak on the FFT spectrum, which may explain the singular frequency of 3.5kT observed in previous torque and PDO measurement[5].

When the field is tilted away from $c$ axis, the peak profile of FFT spectrum in our measurements shows clear consistency with other reports. This indicates that the observation of these four oscillation frequencies may also require the precise alignment of the magnetic field with the crystallographic $c$-direction, which is easily achievable in microstructure samples.

C. Resistivity anisotropy calculation

To confirm that the resistivity anisotropy we observed at low temperature can be qualitatively explained by the quasi-2D Fermi surfaces of $UTe_2$, we calculated the conductivities $\sigma_{ii}$ along three principal axes based on the Fermi surface proposed by ref.9, which is shown in Fig. S1. Within relaxation time approximation, the electrical conductivity tensor is given by[7]:

$$\sigma_{ii} = \int \frac{d\mathbf{k}}{4\pi^3} \sum_n \tau_n(\mathbf{k}) v_{i,n}(\mathbf{k}) v_{i,n}(\mathbf{k}) \left(-\frac{\partial f}{\partial \varepsilon}\right)_{\varepsilon=\varepsilon_n(\mathbf{k})}$$

where $n$ is the band index, and $\tau_n$ is the relaxation time for $n$th band. $v_{i,n}(\mathbf{k})$ is the $i$th component of the quasiparticle group velocity with momentum $\mathbf{k}$ in the $n$th band:

$$v_{i,n}(\mathbf{k}) = \frac{\partial \varepsilon_n(\mathbf{k})}{\hbar \partial k_i}$$

Using the $T = 0$ approximation, where $-\frac{\partial f}{\partial \varepsilon} = \delta(\varepsilon - \varepsilon_F)$, we obtain

$$\sigma_{ii} = \frac{1}{4\pi^3} \sum_n \oint \frac{\tau_n(\mathbf{k}_f) v_{i,n}^2(\mathbf{k}_f)}{|v_{i,n}(\mathbf{k}_f)|} dS$$

Given the difficulty of accounting for electronic correlation, we only consider the geometric anisotropy encoded in the corrugated cylindrical Fermi surface, assuming a constant-valued Fermi velocity $v_{i,n}(\mathbf{k}_f)$. This means that, in this calculation, the Fermi velocity $v_{i,n}(\mathbf{k}_f)$ is replaced by the unit normal vector of the Fermi surface:

$$v_{i,n}(\mathbf{k}_f) = n_{i,n}(\mathbf{k}_f)$$

By further assuming an isotropic and constant relaxation time $\tau_n(\mathbf{k}_f)$, the calculated conductivity tensor reflects purely the anisotropy inherent in the geometry of a quasi-2D Fermi surface. The anisotropy of $\rho_c/\rho_{a,b}$ can then be expressed as:

$$\frac{\rho_c}{\rho_{a,b}} = \frac{\sigma_{a,b}}{\sigma_c}$$

With the above approximations, the calculation gives the following resistivity ratio:

$$\rho_c : \rho_a : \rho_b = 14.35 : 1.57 : 1$$

This result exhibits anisotropy similar to that observed in our microbar measurement. First, the calculated in-plane anisotropy $\frac{\rho_a}{\rho_b} = 1.57$ agrees well with the experimental value $\frac{\rho_a}{\rho_b} \approx 2$. Given that we assumed only a simple form of scattering, this further supports our conclusion that the previously observed results where $\rho_a < \rho_b$, are likely influenced by enhanced disorder scattering, as evidenced by the significantly higher residual resistivity[8]. Second, the calculated anisotropy $\frac{\rho_c}{\rho_b}$ is the highest, as expected for the cylindrical shape of the Fermi surface. Its value underestimates the experimental anisotropy by a factor of 3, which is not surprising given the purely geometric input of our electronic model in the absence of correlated behavior. Evidently, the anisotropy observed in our transport measurement is consistent with the quasi-2D Fermi surface picture supported by QO and ARPES experiments.

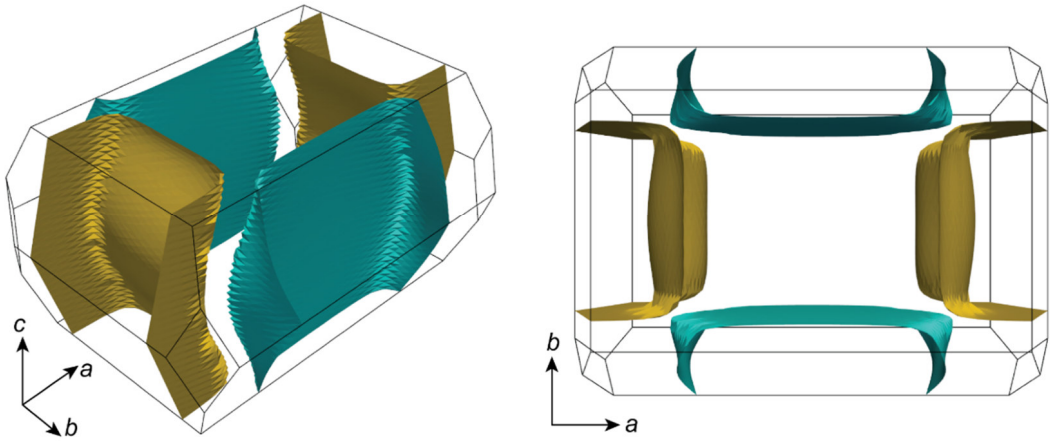

Fig.S1. Fermi surface of $UTe_2$ from the tight-binding calculation in ref.9

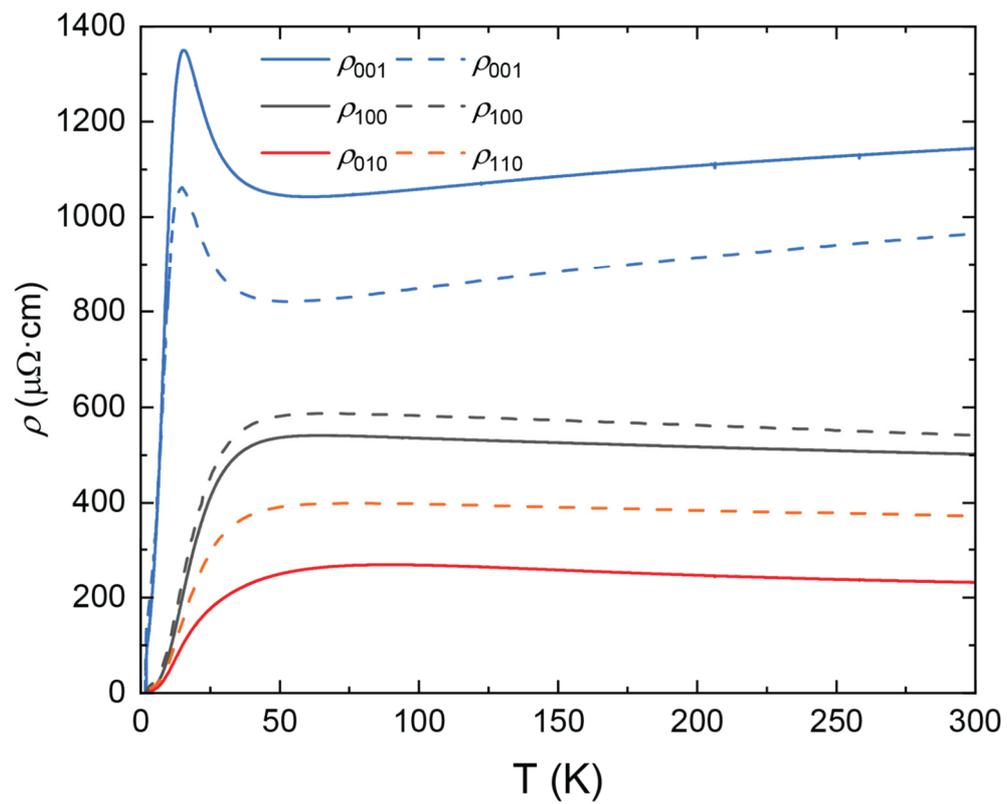

Fig.S2. Comparison of resistivity measured from microbar(solid curves) and bulk single crystal(dashed curves) grown by molten salt flux method. Data of bulk resistivity comes from Ref.3

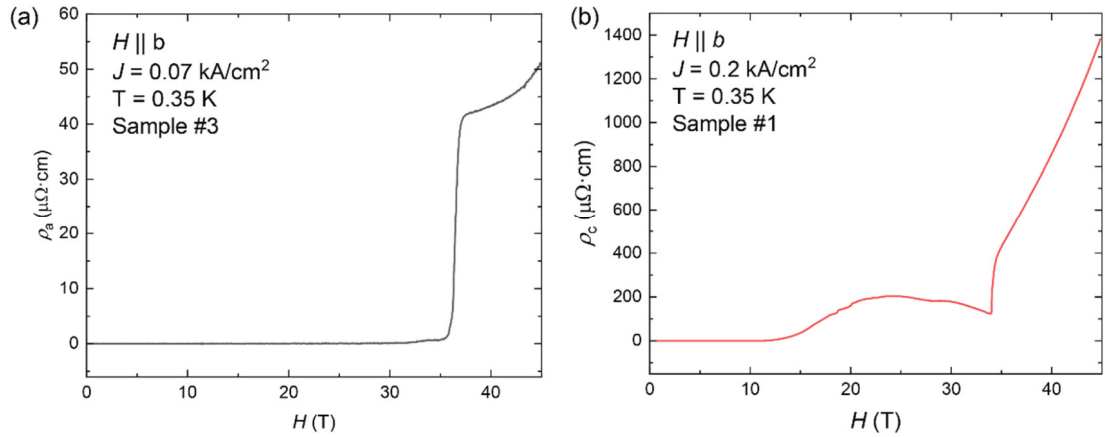

Fig.S3. Resistivity along *a*-and *c*-axis with magnetic field applied along *b* axis from 0 to 45 T.